\documentclass[twocolumn,aps,prl,showpacs,floatfix]{revtex4}
\usepackage{graphicx}
\usepackage{times}
\usepackage{nicefrac}
\usepackage{amsmath}
\usepackage{amsfonts}
\usepackage{amssymb}
\usepackage{amsthm}
\usepackage{epsf}
\usepackage{bm}
\usepackage{times}

\usepackage{dcolumn}

\newcolumntype{.}{D{x}{}{-1}}

\newcommand{\bfr}{{\bf r}}
\newcommand{\bfx}{{\bf x}}

\newcommand{\bfp}{{\bf p}}
\newcommand{\bfA}{{\bf A}}
\newcommand{\bfD}{{\bf D}}
\newcommand{\balpha}{{\mbox{\boldmath$\alpha$}}}
\newcommand{\bnabla}{{\mbox{\boldmath$\nabla$}}}

\newcommand{\be}{\begin{eqnarray}}
\newcommand{\ee}{\end{eqnarray}}
\newcommand{\la}{\langle}
\newcommand{\ra}{\rangle}

\newcommand{\veps}{\varepsilon}

\begin{document}

\title{Recoil effect on the $g$ factor of Li-like ions}

\author{V. M. Shabaev}

\affiliation {Department of Physics, St.Petersburg State University,
Universitetskaya 7/9,
199034 St.Petersburg, Russia}

\author{D. A. Glazov}

\affiliation {Department of Physics, St.Petersburg State University,
Universitetskaya 7/9,
199034 St.Petersburg, Russia}

\author{A. V. Malyshev}

\affiliation {Department of Physics, St.Petersburg State University,
Universitetskaya 7/9,
199034 St.Petersburg, Russia}

\author{I. I. Tupitsyn}

\affiliation {Department of Physics, St.Petersburg State University,
Universitetskaya 7/9,
199034 St.Petersburg, Russia}

\begin{abstract}

  The nuclear recoil effect on the $g$ factor of Li-like ions is evaluated.
  The one-electron recoil contribution is treated
  within the framework of
  the rigorous QED approach to
  first order in the electron-to-nucleus mass ratio
  $m/M$ and to all orders in the parameter $\alpha Z$.
  These calculations are performed in a range $Z=3-92$.
  The two-electron
  recoil term is calculated for low- and middle-$Z$ ions  within the Breit
  approximation
  using a four-component approach. The results for the two-electron recoil part obtained in the paper
  strongly disagree with the previous calculations performed using
  an effective two-component Hamiltonian.
  The obtained value for the
  recoil effect is used to calculate the isotope shift   of the $g$ factor of Li-like
  $^{A}$Ca$^{17+}$ with $A=40$ and $A=48$ which was recently measured. It is found that the new
  theoretical value for the isotope shift is closer
to the experimental one
  than the  previously obtained value.

\end{abstract}
\maketitle


High-precision measurements of the $g$ factor of highly charged ions
\cite{haf00,ver04,stu11,wag13,lin13,stu13,stu14,koel16}
have triggered a great interest to the corresponding theoretical calculations
\cite{blu97,per97,bei00a,bei00b,sha01,sha02b,nef02,yer02,gla02,gla04,lee05,pac05,jen09,vol14,sha15,cza16,yer17a,yer17b}.
To date, these experiment and theory allowed the most stringent tests
of bound-state quantum electrodynamics (QED) in presence of a magnetic field and provided
the most precise determination of the electron mass \cite{stu14,zat17}.
In Ref. \cite{koel16}  the isotope shift of the $g$ factor of Li-like
$^{A}$Ca$^{17+}$ with $A=40$ and $A=48$ has been measured.

The theoretical value of the $g$-factor isotope shift  
is generally given by a sum of the nuclear recoil (mass shift) and nuclear size
(field shift) contributions. For low- and middle-Z ions it is 
mainly determined by the mass shift, which in the case of the $s$ states is
of pure relativistic origin. The fully relativistic theory of the nuclear recoil
effect can be formulated only in the framework of QED. 
Moreover,  the mass shift is the only effect which requires the employment of
the bound-state QED
theory  beyond the external field approximation, providing a unique access to QED beyond
the Furry picture at strong-coupling regime \cite{koel16}.  

In Ref. \cite{koel16} the theoretical value for the $g$-factor mass shift
of  Li-like calcium was obtained combining the calculations of the
one-electron recoil contribution to all orders in $\alpha Z$
and the two-electron recoil contribution within the Breit approximation.
While the one-electron contribution was directly evaluated using the QED theory
 \cite{sha01,sha02a}, the two-electron part was obtained by extrapolating
the lowest-order relativistic results from
Refs. \cite{yan01,yan02}.
Combined with the nuclear size effect, whose calculation causes no problem,
the theoretical prediction for the isotope shift of the $g$ factor of $^{A}$Ca$^{17+}$ with $A=40$ and $A=48$
was found to be in agreement with the experimental one but 
at the edge of the experimental error bar.

In the present paper we perform the most accurate to-date
evaluation of the nuclear recoil contribution to the $g$ factor
of highly charged Li-like ions. First, we improve in accuracy
the calculation of the one-electron QED recoil contribution for Li-like calcium \cite{koel16}
and extend it to a wide range of the nuclear charge number $Z=3-92$.
Second, we calculate the  two-electron recoil contribution to the $g$ factor
in a range  $Z=3-20$  within the Breit approximation
using a four-component approach 
and investigate reasons for a strong disagreement between
the obtained results and the previous calculations \cite{yan01,yan02}.
Finally, we present the theoretical
prediction for the isotope shift  of the $g$ factor of Li-like
$^{A}$Ca$^{17+}$ with $A=40$ and $A=48$, which also includes
 the nuclear size effect,
and compare it  with the
 experiment \cite{koel16}.
%


The QED theory for the nuclear recoil effect on the 
atomic $g$ factor to first order in the electron to nucleus mass ratio
$m/M$ and to all orders in $\alpha Z$ was developed in Ref. \cite{sha01}.
This theory was employed to derive a complete  $\alpha Z$-dependent formula for the recoil
effect on the $g$ factor of a H-like ion.
The obtained formula can be also applied to a many-electron ion (atom)
with one electron over closed shells, provided
the electron propagators are defined for the vacuum including the closed shells
 \cite{sha02a}. In this case, the formula also
incorporates the two-electron
nuclear recoil contributions to zeroth order in $1/Z$.

We consider an ion  with one electron
over closed shells
which is put into the classical
homogeneous magnetic field, ${\bf A}_{\rm cl}(\bfr)=[{\bf {\cal H}}\times {\bfr}]/2$.
For simplicity, we assume that $ {\bf {\cal H}}$ is directed along the $z$ axis.
According to Refs.  \cite{sha01,sha03},
to zeroth order in $1/Z$, the $m/M$
nuclear recoil contribution to the $g$ factor for a  state $a$
is given by 
($\hbar=c=1$, $e<0$)
\begin{eqnarray} \label{rec_tot}
\Delta g&=&\frac{1}{\mu_0 m_a}\frac{i}{2\pi M}
\int_{-\infty}^{\infty} d\omega\;
\Biggl[\frac{\partial}{\partial {\cal H}}
\langle \tilde{a}|[p^k-D^k(\omega)+eA_{\rm cl}^k]
\nonumber\\
&&\times\tilde{G}(\omega+\tilde{\veps}_a)
[p^k-D^k(\omega)+eA_{\rm cl}^k]
|\tilde{a}\rangle
\Biggr]_{{\cal H}=0}\,.
\label{06recoilt}
\end{eqnarray}
Here $\mu_0$ is the Bohr magneton, $m_a$ is the angular momentum
projection of the state under consideration,
 $p^k=-i\nabla^k$ is the momentum operator, 
$D^k(\omega)=-4\pi\alpha Z\alpha^l D^{lk}(\omega)$,
\begin{eqnarray} \label{06photon}
D^{il}(\omega,{\bf r})&=&-\frac{1}{4\pi}\Bigl\{\frac
{\exp{(i|\omega|r)}}{r}\delta_{il}\nonumber\\
&&+\nabla^{i}\nabla^{l}
\frac{(\exp{(i|\omega|r)}
-1)}{\omega^{2}r}\Bigr\}\,
\end{eqnarray}
is the transverse part of the photon propagator in the Coulomb 
gauge,  $\balpha$ is a vector
incorporating the Dirac matrices, and
the summation over the repeated indices is implicit.  
The tilde sign indicates that the corresponding quantity
(the wave function, the energy, and the Coulomb Green function
$\tilde{G}(\omega)$)
must be calculated in presence of the
magnetic field. 
Since we consider an ion with one
valence electron over the closed shells,   
 the Coulomb Green function is defined as
$\tilde{G}(\omega)=\sum_{\tilde{n}}|\tilde{n}\rangle \langle 
\tilde{n}|[\omega-\tilde{\veps}_n +  i\eta(\tilde{\veps}_n - \tilde{\veps}_{\rm F})]^{-1}$,
where $ \tilde{\veps}_{\rm F} $ is the Fermi energy and $\eta \rightarrow 0$.  
Formula  (\ref{06recoilt}) includes 
 both one- and two-electron
 nuclear recoil contributions to zeroth order in $1/Z$.
For the $(1s)^22s$ state of a Li-like ion, the $(1/Z)^0$ two-electron
contribution is equal to zero. However, this formula can be used
to derive an effective two-electron recoil operator which describes
the recoil effect on the $g$ factor within the Breit approximation.
The expression for this operator is given below.

First, we consider the one-electron contribution. 
For the practical calculations, it
is conveniently
represented by a sum of low-order
and  higher-order
terms,
$\Delta g=\Delta g_{\rm L}+\Delta g_{\rm H}$, where 

\be \label{low}
\Delta g_{\rm L}&=&\frac{1}{\mu_0 {\cal H}  m_a}
\frac{1}{M} \la \delta a|
\Bigr[ \bfp^2
 -\frac{\alpha Z}{r}\bigr(\balpha+\frac{(\balpha\cdot\bfr)\bfr}{r^2}\bigr)
  \cdot\bfp
  \Bigr]|a\ra\nonumber\\
&& -\frac{1}{ m_a}\frac{m}{M}\la
a|\Bigl([\bfr \times \bfp]_z -\frac{\alpha Z}{2r}[\bfr \times \balpha ]_z\Bigr)|a\ra\,,
\ee
\be  \label{high}
\Delta g_{\rm H}&=&\frac{1}{\mu_0  {\cal H} m_a}
\frac{i}{2\pi M} \int_{-\infty}^{\infty} d\omega\;\Bigl\{ \la
\delta a|\Bigl(D^k(\omega)-\frac{[p^k,V]}{\omega+i0}\Bigr)\nonumber\\
&&\times G(\omega+\veps_a)\Bigl(D^k(\omega)+\frac{[p^k,V]}{\omega+i0}\Bigr)|a\ra \nonumber\\
&&+\la a|\Bigl(D^k(\omega)-\frac{[p^k,V]}{\omega+i0}\Bigr)
G(\omega+\veps_a)\nonumber\\
&&\times\Bigl(D^k(\omega)+\frac{[p^k,V]}{\omega+i0}\Bigr) |\delta
a\ra\nonumber\\ &&+ \la a|\Bigl(D^k(\omega)-\frac{[p^k,V]}{\omega+i0}\Bigr)
G(\omega+\veps_a)(\delta V-\delta \veps_a)\nonumber\\ &&\times G(\omega+\veps_a)
\Bigl(D^k(\omega)+\frac{[p^k,V]}{\omega+i0}\Bigr)|a\ra \Bigr\}\,. \label{eq3}
\ee
Here
 $V(r)=-\alpha Z/r$ is the Coulomb
potential of the nucleus,
 $\delta V(\bfx)=-e\balpha \cdot\bfA_{\rm cl}(\bfx)$,
$G(\omega)=\sum_n|n\ra \la n|[\omega-\veps_n(1-i0)]^{-1}$ is the Dirac-Coulomb Green
function, $\delta \veps_a=\la a|\delta V|a\ra$, and $|\delta a\ra=\sum_n^{\veps_n\ne
  \veps_a}|n\ra\la n|\delta V|a\ra (\veps_a-\veps_n)^{-1}$.
The low-order term can be derived from the relativistic
Breit equation, while the derivation of the higher-order 
term requires the employment of QED beyond the Breit approximation.
For this reason, we term them as non-QED
and QED one-electron contributions, respectively.

To derive the effective two-electron recoil operator we need to
consider  in Eq. (\ref{rec_tot}) the two-electron contributions 
which describe the interaction of the valence electron with
the closed shell electrons. It can easily be done according to
the corresponding prescriptions in Refs. \cite{sha02a,shch15}. Within
the Breit approximation,  we obtain
\begin{eqnarray}
\Delta g_{\rm int}= \Delta g_{\rm int}^{(1)}+\Delta g_{\rm int}^{(2)},
\end{eqnarray}
where
\be \label{g2_1}
 \Delta g_{\rm int}^{(1)}&=&-\frac{2}{\mu_0 {\cal H}  m_a}
\frac{1}{M} \sum_c \Bigl\{ \la a|p^i|c\ra \la c|p^i| \delta a\ra\nonumber\\
 &&-\la a|p^i|c\ra \la c|D^i| \delta a\ra-\la a|D^i|c\ra \la c|p^i| \delta a\ra\nonumber\\
&&+\la a|p^i| \delta c \ra \la c|p^i|a\ra-\la a|p^i| \delta c \ra \la c|D^i|a\ra\nonumber\\
&&-\la a|D^i| \delta c \ra \la c|p^i|a\ra\Bigr\}\,,\\
\label{g2_2}
 \Delta g_{\rm int}^{(2)}&=&-\frac{1}{ m_a}
\frac{m}{M}\epsilon_{3kl} \sum_c \Bigl\{\la a|x^l|c\ra \nonumber\\
&&\times\la c|(p^k-D^k)| a\ra\nonumber\\
&&+\la a|(p^k-D^k)| c\ra \la c|x^l|a\ra\Bigr\}\,.
\ee
Here
\be
\bfD\equiv \bfD(0)=\frac{\alpha Z}{2r}\bigl(\balpha
+\frac{(\balpha\cdot\bfr)\bfr}{r^2}\bigr)\,,  
\ee
$\epsilon_{ikl}$ is the Levi-Civita symbol,
$|\delta c\ra=\sum_n^{\veps_n\ne
  \veps_c}|n\ra\la n|\delta V|c\ra (\veps_c-\veps_n)^{-1}$, and the summation
($c$) runs over the closed shells. The $\Delta g_{\rm int}^{(1)}$ term
corresponds to the combined interaction due to $\delta V$ and 
the two-electron part of the effective recoil Hamiltonian
(see Ref. \cite{sha98} and references therein):
\be \label{br1}
H_M=\frac{1}{2M}\sum_{i,k}\Bigl[\bfp_i\cdot \bfp_k
  -\frac{\alpha Z}{r_i}\bigr(\balpha_i+\frac{(\balpha_i\cdot\bfr_i)\bfr_i}{r_i^2}\bigr)
  \cdot\bfp_k\Bigr]\,.
\ee
The one-electron part of this operator corresponds to the first term
in Eq. (\ref{low}).
The $\Delta g_{\rm int}^{(2)}$ term
leads to the following magnetic recoil 
operator:
\be \label{br2}
H_M^{\rm magn}&=&-\mu_0 {\cal H}
\frac{m}{M}
\sum_{i, k}\Bigl\{[\bfr_i\times \bfp_k]\nonumber\\
&&-\frac{\alpha Z}{2r_k}\Bigl[\bfr_i\times\Bigl(\balpha_k
   +\frac{(\balpha_k\cdot\bfr_k)\bfr_k}{r_k^2}\Bigr)\Bigr]
  \Bigr\}\,,
\ee
where we have added the corresponding one-electron part from  Eq. (\ref{low}).
The first term in the right-hand side of Eq. (\ref{br2}) 
defines the nonrelativistic contribution derived previously by Phillips \cite{phi49}.

Thus, within the lowest-order relativistic (Breit) approximation the recoil 
effect on the $g$ factor to the first order in $m/M$ can be evaluated
by adding the operators (\ref{br1}) and (\ref{br2})
to the Dirac-Coulomb-Breit Hamiltonian, considered in the
presence of the external magnetic field. As mentioned above,
for the $(1s)^22s$ state of a Li-like ion the $(1/Z)^0$ two-electron
recoil contribution equals zero. However, we can use the derived effective operators
to evaluate the  $1/Z$ and higher-order contributions to the recoil effect within the
Breit approximation.


For a point-charge nucleus, 
the low-order one-electron term $\Delta g_{\rm L}$ can be evaluated
analytically \cite{sha01}:
\begin{eqnarray} \label{06shabaeveq112}
\Delta g_{\rm L}=-\frac{m}{M}
\frac{2\kappa^2\veps^2+\kappa m \veps-m^2}{2m^2j(j+1)}\,,
\end{eqnarray}
where $\veps$ is the Dirac energy and $\kappa=(-1)^{j+l+1/2}(j+1/2)$ is the
angular momentum-parity quantum number.
To the leading orders in $\alpha Z$, we have
\begin{eqnarray} \label{06shabaeveq113}
\Delta g_{\rm L}&=&-\frac{m}{M}\,
\frac{1}{j(j+1)}\Bigl[\kappa^2+\frac{\kappa}{2}
  -\frac{1}{2}\nonumber\\
 && -\Bigl(\kappa^2+\frac{\kappa}{4}\Bigr)
\frac{(\alpha Z)^2}{n^2} + \cdots \Bigr]\,.
\end{eqnarray}
It can be seen that for an $s$ state ($\kappa=-1$)
the non-relativistic contribution to  $\Delta g_{\rm L}$
is equal to zero and, therefore, the low-order term is of pure relativistic  
($\sim (\alpha Z)^2$) origin.

The numerical calculation of the higher-order one-electron contribution (\ref{high})
was performed in the same way as 
in Refs. \cite{sha02b,yer16}.
After the integration over angles, the summation over the intermediate electron states
was carried out using the finite basis set method with
the basis function constructed from B-splines \cite{sap96}. The $\omega$ integration
was performed analytically for the simplest ``Coulomb'' contribution (the term without
 the $\bfD$ vector) 
and numerically
for the ``one-transverse'' and ``two-transverse'' photon contributions
(the terms with one and two $\bfD$ vectors, respectively)
using the standard
Wick's rotation.
The higher-order (QED) contribution
$\Delta g_{\rm H}$ for the $2s$ state
is conveniently
expressed in terms of the function $P^{(2s)}(\alpha Z)$,
\be
\label{P}
\Delta g_{\rm H}^{(2s)}=\frac{m}{M}\frac{(\alpha Z)^5}{8} P^{(2s)}(\alpha Z)\,.
\ee
The corresponding numerical results are presented in  Table~\ref{H-like}.
The uncertainties have been obtained by studying the stability
of the results with respect to a change of the basis set size.
For $Z=20$ the presented result agrees with that from Ref. \cite{koel16}
but is given to a higher accuracy.

To get the total one-electron recoil contribution, we should also account
for the radiative ($\sim\alpha$)  and second-order (in $m/M$) recoil corrections.
To the lowest order in $\alpha Z$ these corrections were evaluated 
in Refs. \cite{gro71,clo71,pac08,eid10}.

As noted above, for the $(1s)^22s$ state of a Li-like ion the two-electron
recoil contribution to the $g$ factor is equal zero, if one neglects the
interaction between the electrons. This approximation 
 corresponds to zeroth order in $1/Z$.
The recoil contributions of the first and higher orders
in $1/Z$ have been evaluated within the Breit approximation using the
operators (\ref{br1}), (\ref{br2}) and the standard expression for
the Dirac-Coulomb-Breit Hamiltonian:
\begin{eqnarray} \label{dcb}
H^{\rm DCB}=
\Lambda^{(+)}\Bigl[\sum_{i}h_i^{\rm D}  +\sum_{i<k}V_{ik}\Bigr]\Lambda^{(+)}\,,
\end{eqnarray}
where the indices $i$ and $k$ enumerate the atomic electrons,
$\Lambda^{(+)}$ is the product of the one-electron
projectors on the positive-energy states
(which correspond to the potential $V+\delta V$,
where $V$ is the Coulomb potential of the nucleus and $\delta V$
describes the interaction with the external magnetic field),
$h_i^{\rm D}$ is the one-electron Dirac Hamiltonian
including $\delta V$,
\begin{eqnarray} \label{br}
V_{ik} &=& e^2\alpha_i^{\rho}\alpha_k^{\sigma}D_{\rho \sigma}(0,\bfr_{ik})
= V^{\rm C}_{ik}+ V^{\rm B}_{ik}\nonumber \\
&=&\frac{\displaystyle \alpha}{\displaystyle r_{ik}} 
-\alpha\Bigl[\frac{\displaystyle{ {\balpha}_i\cdot {\balpha}_k}}
{\displaystyle{ r_{ik}}}+\frac{\displaystyle 1}{\displaystyle 2}
( {\balpha}_i\cdot{\bnabla}_i)({ {\balpha}_k\cdot\bnabla}_k)
r_{ik}\Bigr]
\,
\end{eqnarray}
is the sum of the Coulomb and Breit electron-electron interaction operators.

Let us consider first the calculation of the $1/Z$ recoil contribution, which
can be evaluated using perturbation theory.
This contribution is conveniently represented by a sum of four terms:
\be \label{g_ind}
\Delta g_{\rm int}^{(1/Z)} = \Delta  g_{\rm int}^{\rm (1)}
+ \Delta g_{\rm int}^{\rm (2)}
+  \Delta g_{\rm int}^{\rm (1m)}
+ \Delta g_{\rm int}^{\rm (2m)}\,,
\ee
where $\Delta  g_{\rm int}^{\rm (1)}$ combines the one-electron
non-magnetic recoil term from Eq. (\ref{br1})
with the electron-electron
interaction (\ref{br})
and with the magnetic interaction $\delta V$,
 $\Delta  g_{\rm int}^{\rm (2)}$ combines the two-electron
 non-magnetic recoil term from Eq. (\ref{br1})
 with the electron-electron
 interaction (\ref{br})
 and with the magnetic interaction $\delta V$, 
 $\Delta  g_{\rm int}^{\rm (1m)}$ combines the one-electron
magnetic recoil term from Eq. (\ref{br2})
with the electron-electron
interaction (\ref{br}), and
 $\Delta  g_{\rm int}^{\rm (2m)}$ combines the two-electron
magnetic recoil term from Eq. (\ref{br2})
with the electron-electron
interaction (\ref{br}). The numerical evaluation of all these terms
has been performed for extended nuclei in the range $Z=3-20$
using
the finite basis set method with the basis functions
constructed from B-splines.
The results, which are expressed in terms of the function $B(\alpha)$ defined
by
\be \label{B}
\Delta g_{\rm int}^{(1/Z)}=\frac{m}{M}\frac{(\alpha Z)^2}{Z} B(\alpha Z)\,,
\ee
are presented in Table~\ref{Li-like}.
All digits presented in the table should be correct.
The extrapolation to the limit $\alpha Z\rightarrow 0$ leads to $B(0)=-0.5155(2)$.
This value disagrees with the corresponding coefficient
 $B(0)=-0.8603(8)$
which can be derived (see  Ref. \cite{yer17b})
by fitting the lowest-order
relativistic results
of the fully correlated calculations within 
the framework of a two-component approach
performed 
by Yan \cite{yan01,yan02}.
To find out the reasons for this disagreement, we have also
evaluated the $1/Z$ recoil corrections using the effective two-component
Hamiltonian approach
\cite{heg73,heg75,yan01,pac08,wien14}.
The calculations have been performed by perturbation theory starting  
with the nonrelativistic independent-electron approximation. 
The summations over  electron spectra have been carried out 
using the finite basis set method for the Schr\"odinger equation
with the basis functions constructed from B splines \cite{sap96}.
With this approach, we obtain
$B(0)=-0.8603$, provided we account for the same 
contributions  as described in  Refs. \cite{heg75,yan01,yan02}.
This corresponds to the evaluation of the spin-dependent terms
in the magnetic-field dependent part of the effective
two-component Hamiltonian 
with the Schr\"odinger wave function.
In the previous calculations \cite{heg75,yan01,yan02}
it was assumed that only these terms
contribute for the $s$ states. Our study
showed, however, that this is not the case.
We have found that there exist some additional
contributions
to the lowest relativistic order.
To the first order in $1/Z$,
these contributions originate from the spin-independent terms
in the magnetic-field dependent part of the
effective Hamiltonian
(the first term in Eq. (\ref{br2})) if they are
combined with
the spin-orbit and spin-other-orbit coupling terms
in the non-magnetic part of the  two-component Hamiltonian
(the expressions for
these couplings see, e.g., in Ref. \cite{heg73}).
The spin-orbit coupling leads to a nonzero result if it is
combined with the Coulomb electron-electron interaction.
The evaluation
of these terms gives additionally 0.3447 to $B(0)$. This leads to the total result
$B(0)=-0.5156$ which agrees with the value obtained in our four-component
approach.

The evaluation of the second and higher-order in $1/Z$ contributions within
the Breit approximation was also based on the
operators (\ref{br1}), (\ref{br2}) and the standard expression for
the Dirac-Coulomb-Breit Hamiltonian (\ref{dcb}). This was done
by the use of
a recently developed recursive perturbative approach \cite{gla17,mal17}.
The results, which are expressed in terms of the function $C(\alpha Z)$ defined
by
\be \label{C}
\Delta g_{\rm int}^{(1/Z^{2+})}=\frac{m}{M}\frac{(\alpha Z)^2}{Z^2} C(\alpha Z)\,,
\ee
are presented in Table~\ref{Li-like-2}. The $C^{\rm (1+2)}(\alpha Z)$
and  $C^{\rm (1m+2m)}(\alpha Z)$ parts, presented in the table, correspond to the non-magnetic
and magnetic recoil contributions defined by operators  (\ref{br1}) and (\ref{br2}),
respectively. The indicated error bars are due to the numerical
uncertainties of the computation.

To derive the total
value of the isotope shift, we need also to  evaluate the nuclear size
effect. In case of Ca isotopes, this contribution
can be calculated in the one-electron approximation using
the analytical formula from Ref. \cite{gla02}. The root-mean-square 
nuclear charge radii and the related uncertainties
were taken from Ref. \cite{ang13}.

The individual contributions to the isotope shift of the $g$ factor
for $^{40}$Ca$^{19+}$ - $^{48}$Ca$^{19+}$ are presented in
Table~\ref{isotope_shift}.
The uncertainty of the finite nuclear size contribution
includes both the nuclear radius and shape variation effects.
The shape variation uncertainty was estimated as a difference between 
the calculations performed for the Fermi and sphere nuclear models.
The total theoretical value of the isotope
shift amounts  to   $\Delta g_{\rm IS}^{\rm (theor)} = 11.056(16)\times 10^{-9}$. This value
differs from its previous evaluation, 
$\Delta g_{\rm IS}^{\rm (theor)} = 
10.305(27)\times 10^{-9}$  \cite{koel16},
which included
the two-electron recoil contribution obtained by extrapolating the corresponding
results from
Refs. \cite{yan01,yan02}, and is significantly closer  to  the experimental
value, $ \Delta g_{\rm IS}^{\rm (exp)} = 11.70(1.39)\times 10^{-9}$   \cite{koel16}.

Concluding, 
in this paper we have evaluated the nuclear recoil effect on the
$g$ factor of Li-like ions. The calculations included the $m/M$ one-electron
recoil correction in the framework of the fully relativistic formalism  and
the two-electron recoil contribution within the Breit approximation. A large
discrepancy was found between the present result for the two-electron recoil
contribution obtained using the four-component approach within the
Breit approximation and its previous calculation performed using the
effective two-component Hamiltonian. An analysis of the discrepancy
showed that some important contributions were omitted in the previous
works. As the result, the most precise to-date theoretical values for the
recoil effect on the $g$ factor of Li-like ions have been obtained.
Combining the nuclear recoil and size effects, the isotope shift of
the $g$ factor of Li-like $^{A}$Ca$^{17+}$ with $A=40$ and $A=48$
has been evaluated providing better agreement between theory
and experiment.
We hope that the obtained results will also pave the way
for QED tests beyond the Furry picture in experiments with highly charged
ions  which are planned at the Max-Planck-Institut f\"ur Kernphysik in Heidelberg
and at the HITRAP/FAIR facilities in Darmstadt.


We thank Krzysztof Pachucki and Vladimir Yerokhin for valuable discussions.
This work was supported by the Russian Science Foundation (Grant No. 17-12-01097).
%
%

\begin{table}
\caption{The higher-order (QED) recoil contribution to the $2s$ $g$ factor,
expressed in terms of the function $P^{(2s)}(\alpha Z)$ defined by
Eq. (\ref{P}).}
\label{H-like}
\begin{tabular}{cr@{.}lr@{.}lr@{.}lr@{.}l} \hline
$Z$& \multicolumn{2}{c}{$P^{(2s)}_{\rm Coul}$}
                    & \multicolumn{2}{c}{$P^{(2s)}_{\rm tr1}$}
                                     & \multicolumn{2}{c}{$P^{(2s)}_{\rm tr2}$ }
                                                     & \multicolumn{2}{c}{$P^{(2s)}(\alpha Z)$}
\\
\hline
   3  &   $-$1&082   &   37&719      &   $ -$22&644   &    13&993(1)  \\
 4  &   $-$1&067   &   29&519      &   $ -$15&755   &    12&697(1)  \\
 
   6  &   $-$1&0403   &   21&1043      &   $ -$9&1264   &    10&9376(2)  \\
  8  &   $-$1&0160   &   16&7563      &   $ -$5&9884   &    9&7519(1)  \\  
   10  &   $-$0&9943   &   14&0726      &   $ -$4&2021   &    8&8762(1)  \\
   12  &   $-$0&9749   &  12&2396  &      $-$3&0704      &    8&1943(1)\\
      14  &   $-$0&9575   &   10&9033      &   $ -$2&3010   &    7&6447(1)  \\
16 &        $-$0&9422  &     9&8841 &      $-$1&7509  &   7&1911(1) \\
      18  &   $-$0&9287   &   9&0809      &   $ -$1&3421   &    6&8101(1)  \\
 20  &   $-$0&9169   &   8&4322      &   $ -$1&0292   &    6&4860(1)  \\
  30  &   $-$0&8818   &   6&4789      &   $ -$0&1810   &    5&4160(1)  \\
  40  &   $-$0&8836   &   5&5765      &   0&1911   &    4&8840(1)  \\
   50  &   $-$0&9244   &   5&1802      &   0&4168   &    4&6727(1)  \\
    60  &   $-$1&0150   &   5&1326      &   0&6005   &    4&7182(1)  \\
  70  &   $-$1&181   &   5&426      &   0&795   &    5&040(1)  \\
   80  &   $-$1&482   &   6&186      &   1&050   &    5&753(3)  \\
     90  &   $-$2&07   &   7&82      &   1&45   &    7&20(1)  \\
 92  &   $-$2&26   &   8&33      &   1&57   &    7&64(2)  \\
\hline

\end{tabular}
\end{table}

\begin{table}
  \caption{The $1/Z$ recoil contribution to the $g$ factor of the
    $(1s)^2 2s$ state of
    Li-like ions,
expressed in terms of the function $B(\alpha Z)$ defined by
equation (\ref{B}). The individual contributions correspond to 
the related terms in Eq. (\ref{g_ind}).}
\label{Li-like}
\begin{tabular}{cr@{.}lr@{.}lr@{.}lr@{.}lr@{.}l} \hline
$Z$& \multicolumn{2}{c}{$B^{\rm (1)}(\alpha Z)$}
                    & \multicolumn{2}{c}{$B^{\rm (2)}(\alpha Z)$}
  & \multicolumn{2}{c}{$B^{\rm (1m)}(\alpha Z)$ }
   & \multicolumn{2}{c}{$B^{\rm (2m)}(\alpha Z)$ }
                         & \multicolumn{2}{c}{$B(\alpha Z)$}
\\
\hline
3  &   $-$0&8835  &   0&0213    &   0&0000   &    0&3466 & $-$0&5157 \\
4  &   $-$0&8836  &   0&0213    &   0&0000   &    0&3465 & $-$0&5158 \\
6  &   $-$0&8839  &   0&0214    &   0&0000  &    0&3464 & $-$0&5161 \\
8  &   $-$0&8844  &   0&0216    &   0&0000   &    0&3462 & $-$0&5166 \\
10  &   $-$0&8849  &   0&0218    &   0&0000   &    0&3459 & $-$0&5172 \\
12  &   $-$0&8856  &   0&0220    &   0&0001   &    0&3456 & $-$0&5179 \\
14  &   $-$0&8864  &   0&0223    &  0&0001   &    0&3453 & $-$0&5187 \\
16  &   $-$0&8873  &   0&0227    &  0&0001   &    0&3449 & $-$0&5197 \\
18  &   $-$0&8883  &   0&0231    &   0&0001   &    0&3444 & $-$0&5207 \\
20  &   $-$0&8894   &   0&0235      &  0&0001   &    0&3439 & $-$0&5219 \\
\hline
\end{tabular}
\end{table}

\begin{table}
  \caption{The $1/Z^{2}$ and higher-order recoil contribution to the $g$ factor of the
    $(1s)^2 2s$ state of
    Li-like ions,
expressed in terms of the function $C(\alpha Z)$ defined by
Eq. (\ref{C}). The $C^{\rm (1+2)}(\alpha Z)$
and  $C^{\rm (1m+2m)}(\alpha Z)$ parts correspond to the non-magnetic
and magnetic recoil contributions defined by operators  (\ref{br1}) and (\ref{br2}),
respectively.}
\label{Li-like-2}
\begin{tabular}{cr@{.}lr@{.}lr@{.}l} \hline

  $Z$& \multicolumn{2}{c}{$C^{\rm (1+2)}(\alpha Z)$}
                    & \multicolumn{2}{c}{$C^{\rm (1m+2m)}(\alpha Z)$} 
                       & \multicolumn{2}{c}{$C(\alpha Z)$}
\\
\hline
3  &   0&61(5)  &   $-$0&75(5)   & $-$0&14(7) \\
4  &   0&59(3)  &   $-$0&76(3)    &  $-$0&17(4) \\
6  &   0&566(10)  &  $-$0&775(10) & $-$0&209(14) \\
8  &   0&556(5)  &  $-$0&782(5)    & $-$0&226(7) \\
10  &   0&550(3)  &  $-$0&786(3)   & $-$0&236(4) \\
12  &   0&546(2)  &  $-$0&789(2)   & $-$0&243(3) \\
14  &   0&545(2)  &  $-$0&790(2)   & $-$0&245(3) \\
16  &   0&543(2)  &  $-$0&791(2)   & $-$0&248(3) \\
18  &   0&542(1)  &   $-$0&792(1)  & $-$0&250(2) \\
20  &   0&542(1)   &  $-$0&792(1)      & $-$0&250(2) \\
\hline
\end{tabular}
\end{table}

\begin{table}
\caption{ Isotope shift of the $g$ factor: $^{40}{\rm Ca}^{17+} -\; ^{48}{\rm Ca}^{17+}$, in units $10^{-9}$.}
\label{isotope_shift}
\begin{center}
\begin{tabular}{ll}\hline
  One-electron non-QED nuclear recoil   & $\;\;\; 12.240 $ \\
  Two-electron non-QED nuclear recoil     &  $\;\; - 1. 302$ \\
  QED nuclear recoil: $\sim m/M$ &           $\;\;\;\;\; 0.123 (12) $ \\
  QED nuclear recoil: $\sim \alpha (m/M)$ &           $\;\;-0.009 (1)$ \\
  Finite nuclear size & $\;\;\;\;\; 0.004(11)$ \\
  Total theory & $ \;\;\; 11.056(16)$ \\
Experiment \cite{koel16}   & $ \;\;\; 11.70(1.39)$ \\ \hline
\end{tabular}
\end{center}
\end{table}




\end{document}